# Spatially Resolved Spectroscopic Study of nearby Seyfert Galaxies: Implications for a Population of "Missed" Seyferts at High-z

Junjie Xia[1,3], Matthew A. Malkan[1], Nathaniel R. Ross[1,2], and Agnes J. Ancheta[1]
[1] Department of Physics and Astronomy, University of California, Los Angeles, 430 Portola Plaza, Los Angeles, CA 90095, USA; seanxia8@gmail.com
[2] Raytheon Space and Airborne Systems, 2000 E El Segundo Boulevard, El Segundo, CA 90245, USA

## Abstract

We present mosaicked long-slit spectral maps of 18 nearby Active Galactic Nuclei (AGNs), 2 LINERs, and 4 star-forming galaxies. With the resulting data cubes taken using the Kast dual spectrograph on the 3 m Shane telescope of the Lick Observatory, we measure the aperture effects on the spectroscopic classification of AGNs. With more starlight included in a larger aperture, the nuclear spectrum that is Seyfert-like may become contaminated. We generated standard spectroscopic classification diagrams in different observing apertures. These show quantitatively how the ensemble of Seyferts migrates toward the H II region classification when being observed with increasing aperture sizes. But the effect ranges widely in individual active galaxies. Some of the less luminous Seyferts shift by a large amount, while some others barely move or even shift in different directions. We find that those Seyfert galaxies with the fraction of nuclear H$\alpha$ emission lower than 0.2 of the host galaxy, 2–10 keV hard X-ray luminosity lower than $10^{43}$ erg s$^{-1}$, and the observed nuclear [O III] luminosity lower than $10^{40.5}$ erg s$^{-1}$, are more likely to change activity classification type when the entire host galaxy is included. Overall, 4 of our 24 galaxies (18 Seyferts) change their spectral activity classification type when observed with a very large aperture.

## 1. Introduction

Although most galaxies are now thought to host a central super massive black hole, only a minority of them are currently accreting at sufficient rates to reveal their presence as active galactic nuclei (AGNs). The possibility that observers may have "missed" an active nucleus in some kind of galaxy is a constant concern in this field (e.g., Edelson & Malkan 2012). The usual methods for distinguishing between AGNs and non-AGNs is the spectroscopy of their emission lines. The most definitive AGN signature is the detection of broad wings ($\geqslant 1000$ km s$^{-1}$) of permitted lines (Woo et al. 2008; Bennert et al. 2015). These are produced in the so-called broad-line region in Type 1 AGNs, which—we know from reverberation studies—is within a parsec of the massive black hole and strongly influenced by its gravity (Wandel et al. 1999).

In Type 2 AGNs, these broad lines are not directly observed, possibly because of obscuration, so less direct searches must be made by using the ratios of various narrow emission lines from the so-called narrow-line region (NLR; e.g., Malkan & Oke 1983). To distinguish photoionization from AGNs rather than hot stars, which also produce ionizing photons, narrow emission line ratio diagnostics were developed by Baldwin et al. (1981). The emission line ratios in these widely used "BPT diagrams" are [N II]6583/H$\alpha$, [O III]5007/H$\beta$, [S II]6716,6731/H$\alpha$, [O I]6300/H$\alpha$, and the less standard [O III]5007/[O II]3727 ratios. Using these, Kauffmann et al. (2003) and Kewley et al. (2001, 2006) defined both empirical and theoretical boundaries in BPT diagrams, separating Seyferts, LINERs, and star-forming galaxies. In general, the NLR of Seyferts has stronger high-ionization forbidden line to Balmer line ratios, while the LINERs have weaker [O III]5007/H$\beta$ ratios than Seyferts (Heckman 1980; Ho et al. 1997). On the other hand, the star-forming galaxies have lower ratios of forbidden to Balmer lines.

The high-ionization emission lines of the NLR are powered by high-energy ionizing photons emitted by the accreting black hole in the AGN (Spinoglio & Malkan 1992) in the extreme UV and soft X-rays (Malkan 1988). Their emission is concentrated within the central kiloparsec of the galaxy. However, if the AGN is sufficiently luminous, it can also produce high-ionization NLR emission lines extending out to as much as ∼10 kpc from the galactic center (Villar-Martin et al. 2010; Husemann et al. 2013a). Long-slit spectroscopy shows a positive correlation between the AGN [O III] luminosity and NLR size (Bennert et al. 2002; Greene et al. 2011; Husemann et al. 2013a; Liu et al. 2013a, 2013b; Sun et al. 2018).

Almost all large spectroscopic galaxy surveys have taken spectra with fibers centered on the galaxies' nucleus. Therefore, the observation of Seyfert galaxies at large redshifts tends to include most or all of the host galaxies in the observing aperture. For example, the largest set of spectra, the Sloan Digital Sky Survey (SDSS; York et al. 2000; Eisenstein et al. 2011; Blanton et al. 2017), uses a 3″ fiber diameter, which mixes together an unknown combination of gas ionized by the central AGN, and also by young stars in the disk of the galaxy (Green et al. 2017). As more of the host galaxy light is included, the resulting spectrum will look less like a Seyfert galaxy and more like a star-forming one. Put another way, if a given galaxy spectrum was measured in a fixed slit size, the higher the redshift of the galaxy, the more its measured emission line ratios should shift downward in the BPT

---
[3] Junjie Xia has now moved to The University of Tokyo.

Table 1
12 μm Seyfert Targets Observed

| Galaxy Name (1) | z (2) | Activity Type (3) | Exposure (s) (4) | Date (5) | Weather (6) |
|---|---|---|---|---|---|
| Arp 220 | 0.018126 | Sy 2 | 5400 | 2012 Jun 18 | Clear |
| FSC 01475-0740 | 0.018026 | Sy 2 | 3000 | 2012 Aug 15 | Clear |
| FSC 04385-0828 | 0.015100 | Sy 1 | 3000 | 2012 Dec 7 | Clear |
| IC 5298 | 0.027422 | Sy 2 | 3000 | 2011 Dec 21 | Clear |
| Mrk6 | 0.018813 | Sy 2 | 3600 | 2012 Dec 6 | Clear |
| Mrk 79 | 0.022189 | Sy 1 | 7200 | 2012 Dec 7 | Clear |
| Mrk 463 | 0.050355 | Sy 2 | 4200 | 2012 Jun 20 | Clear |
| Mrk 704 | 0.029234 | Sy 1 | 2400 | 2012 Dec 6 | Clear |
| Mrk 766 | 0.012929 | Sy 1 | 3900 | 2012 Apr 16 | Cirrus |
| Mrk 817 | 0.03145 | Sy 1 | 4800 | 2012 Jun 23 | Cirrus |
| Mrk 1034 | 0.033830 | H II | 4800 | 2012 Dec 6 | Clear |
| NGC 262 | 0.015034 | Sy 2 | 4800 | 2012 Aug 14 | Clear |
| NGC 1056 | 0.005154 | Sa | 5400 | 2012 Aug 16 | Clear |
| NGC 1144 | 0.028847 | Sy 2 | 4800 | 2012 Dec 6 | Clear |
| NGC 1194 | 0.013596 | Sy 1 | 5400 | 2012 Dec 6 | Clear |
| NGC 1241 | 0.013515 | Sy 2 | 9600 | 2012 Dec 7 | Clear |
| NGC 1320 | 0.008883 | Sy 2 | 5400 | 2012 Dec 6 | Clear |
| NGC 1667 | 0.015167 | Sy 2 | 3000 | 2013 Jan 21 | Light Cirrus |
| NGC 1961 | 0.013122 | LINER | 9600 | 2011 Dec 20 | Clear |
| NGC 2639 | 0.011128 | LINER | 6000 | 2012 Dec 6 | Clear |
| NGC 3516 | 0.008836 | Sy 1 | 8400 | 2012 Apr 15 | Cirrus |
| NGC 3982 | 0.003699 | Sy 1 | 9000 | 2012 Apr 15 | Cirrus |
| NGC 5257 | 0.022676 | H II | 6000 | 2011 Dec 20 | Clear |
| NGC 5953 | 0.006555 | H II | 6300 | 2012 Apr 16 | Cirrus |

**Note.** Exposure times refer to the total scan of the red side spectrograph only. The exposure times on the blue and red sides are the same.

diagrams. If the contributions from H II regions become large enough to dominate over the NLR, this aperture effect can cause systematic misclassification. More AGNs at high redshifts would mistakenly be classified as normal star-forming galaxies. This effect has been described by Barger et al. (2001a, 2001b) to explain why 50% of their hard X-ray sources do not have high-ionization signatures of AGNs. Therefore, it is desirable to study the spatial distribution of the emission line ratios from Seyfert nuclei and their host galaxies, in a representative sample.

Recently, surveys of spatially resolved spectroscopy are being made with integral field units (CALIFA, Sanchez et al. 2012; MaNGA, Bundy et al. 2015; SAMI, Croom et al. 2012). They have shown that 2D spectra enable the study with varied aperture sizes on many astronomical parameters, including the spatial distribution of H II and ionizing sources, the star formation rate, the rotation curve probed by neutral hydrogen H I, etc. (Husemann et al. 2013b; Iglesias-Páramo et al. 2013, 2016; Mast et al. 2014; Catalán-Torrecilla et al. 2015; Richards et al. 2016a, 2016b). To study the aperture effect on Seyfert classification, we spatially separated the emission lines coming from the center of the galaxy (produced by its active nucleus) from the emission lines produced in the host galaxy by hot young stars. This allows us to plot separately the line ratios from the two respective contributions in the well-established BPT diagnostic diagrams. Our targets include 18 Seyfert 1 and 2, 2 LINER, and 4 star-forming galaxies. We present our observations in Section 2 and analysis in Section 3. This local sample will help to predict how many Seyfert galaxies may be misclassified at higher redshift, where it is not possible to isolate the nucleus from the host galaxy.

## 2. Observation and Data Reduction

The galaxies we observed, presented in Table 1, are selected from the revised IRAS 12 μm sample of Rush et al. (1993). This contains 893 galaxies in total, with 118 classified as Seyfert 1 or 2. We obtained our data with the Kast double spectrograph on the 3 m Shane telescope of the Lick Observatory. We first aligned a wide slit ($5'' \sim 12''$, various slit widths used throughout the observation period) along the major axis of a target active galaxy nucleus. This position allowed us to extract a $3''$-long by slit-width subaperture from the center of the galaxy that we treated as the nucleus. Then we shifted the slit across the target along its minor axis in both directions with step sizes of $5'' \sim 12''$, until the entire galaxy was "scanned". The use of a wide slit in our observation allowed more light to be collected, reducing the complications caused by poor seeing, even though our observing was usually under good weather conditions. This also sped up the completion of the mosaic maps greatly.

We used the grating No.4 with 1200/5000 grooves/mm blazed at 5000A, which gives a spectral sampling of 0.65 A/pix on the red side. Simultaneously, on the blue side, we used grism No.2, which has 600/4310 grooves/mm blazed at 4310A and 1.02 A/pix. The spectral resolution in the red was found to be 1.95A, and 3.06A in the blue. Hence, even with the wide slits used—we were able to separate most of the lines being used in diagnostic line ratio diagrams, for example, [N II]6583 versus Hα6563.

The average spatial resolution throughout our observation was $2.7(2.68) \pm 0.9(0.86)''$ on the red side and $2.6(2.60) \pm 1.1(1.05)''$ on the blue side. We obtained spatial sampling along the slit to $\sim 3''$ by extracting subapertures. Since the red side spectrograph has $1\,\mathrm{pix} = 0.78''$, while the blue side has

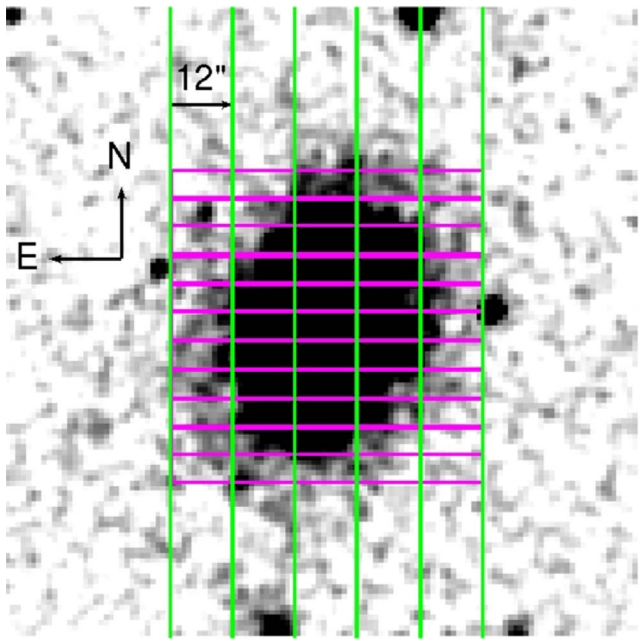

**Figure 1.** Digital Sky Survey image of NGC 1667 with our long slits overlaid in green and extraction apertures in pink. The slit is 12″ wide and 155″ long, running in the north–south direction. Each subaperture is 3″ long. The active nucleus is included in the central subaperture.

1 pix = 0.43″, each 3″ subaperture corresponds to 3.85 pix in the red and 6.98 pix in the blue. We approximated this with subapertures of 4 pix wide in the red and 7 pix wide in the blue. The extracted subapertures are illustrated in Figure 1 with the direct image of NGC 1667 superposed as an example. To get the integrated spectrum of the whole host galaxy, we added up the subapertures out to where the signal-to-noise ratio (S/N) has dropped down to the point, at which the emission lines could not be distinguished from sky background.

The data were reduced using IRAF. We used the Kast He–Hg–Cd arc lamps for wavelength calibration on both the red and blue sides. The flux calibration was done by measuring Feige 15 and Feige 34 as spectrophotometric standard stars. With the extracted subapertures providing spectral information in different locations of a galaxy, we were able to generate an accurate 2D map of emission line fluxes. However, throughout the project we found that once the line ratio got into the star-forming region of BPT diagrams, the classification did not change much with increasing size of the aperture. Hence we only focused on the two most extreme cases—the nucleus spectrum and the integrated spectrum of the entire galaxy. Since the integrated field spectra usually have large components of stellar continuum and hence affect the measurements of the emission lines from the AGNs, we applied the stellar continuum fit pixel-by-pixel with the spectral synthesis code STARLIGHT V.04 (Cid Fernandes et al. 2005) for the integrated fields. In Figure 2, a demonstration of this fit with Mrk 766 is presented. We assumed that the total-aperture continuum is purely stellar, which is not perfectly correct because of the relatively faint nonstellar components in the active nucleus. In overlooking this AGN continuum, especially in the more powerful Seyfert 1 galaxies, our stellar continuum subtraction would tend to overestimate the maximum shifts of the emission line ratios [O III]/H$\beta$ and [N II]/H$\alpha$.

In principle, this same starlight subtraction procedure should have also been applied to the nuclear spectra. It is possible that some of the nuclear spectra have some small contamination from a weak starlight contribution, so this correction would slightly increase the strength of the narrow Balmer emission lines relative to the forbidden lines. In the BPT diagram, this could shift the nuclear spectra slightly down to the left, i.e., slightly closer to the normal/AGN boundary line. However, the practical reality is that the starlight contributions to the nuclear spectra are so small, that they cannot generally be measured with any reliability, at least with spectra of our modest resolution and S/N (Malkan & Filippenko 1983 show the quality of spectra needed to accomplish this accurately). Tests have confirmed that the (uncertain) effects would mostly be negligible. Furthermore, our goal here is to identify the strongest classification changes that can occur when going from nuclear to larger-aperture spectra. We are therefore being deliberately conservative by accepting the emission line measurements in the raw nuclear spectra, uncorrected for the small starlight fraction. In principle, the amount of spectral misclassification might be slightly less than we estimate here, but certainly not more.

In Figure 3, we present the ratio of [N II] to H$\alpha$ at different radii from the galactic center of NGC 1667. The distances to the center in the units of arcsec were derived using the method aforementioned in Figure 1. We can see that the [N II]/H$\alpha$ ratio peaks at the "nucleus" and continuously decreases with growing area enclosed within the aperture. This illustrates the possibility of misclassification in the larger aperture. We also can see that when it approaches to the host galaxy's edge, which is at around 35 arcsec in the case of NGC 1667, the change in the line ratio becomes negligible.

In Figure 4, we present two comparisons between the nuclear and the integrated spectrum. The case of NGC 5953 clearly shows a faster growth of the Balmer lines than the forbidden lines. The emission line ratios in the host galaxy appear larger than those in the nucleus. A counter-example is NGC 3516, for which in both integrated spectra the relative strengths of [N II] and [O III] do not decline, but increase in larger apertures.

This result is similar to what was presented by Moran et al. (2002). They have also observed that some Seyferts would have the relative strength of high-ionization forbidden lines decrease in larger apertures, whereas some others would have enhanced AGN features with increased aperture size. This implies a diversity in the spatial change of emission line ratios. However, most of the broad wings of Seyfert 1 nuclei in our samples were still visible in the integrated spectra, as in NGC 3516. Thus the majority of broad-line objects can hardly be misclassified with good spectroscopy.

## 3. Analysis

### 3.1. Emission Line Ratios

The spectral lines from the nuclei and host galaxies were measured by the SPLOT tool in IRAF after data reduction and starlight subtraction in the larger apertures. We fitted narrow lines with a single Gaussian and decomposed the broad lines with multiple Gaussian components. The line ratios we present in this paper, except [O III]/[O II], are the observed ones with no corrections applied for extinction. This is because our target galaxies are usually at high latitudes, the Milky Way reddening is small (<10%) and would have the equal effect on the nuclear and total apertures. For the [O III]/[O II] extinction correction,

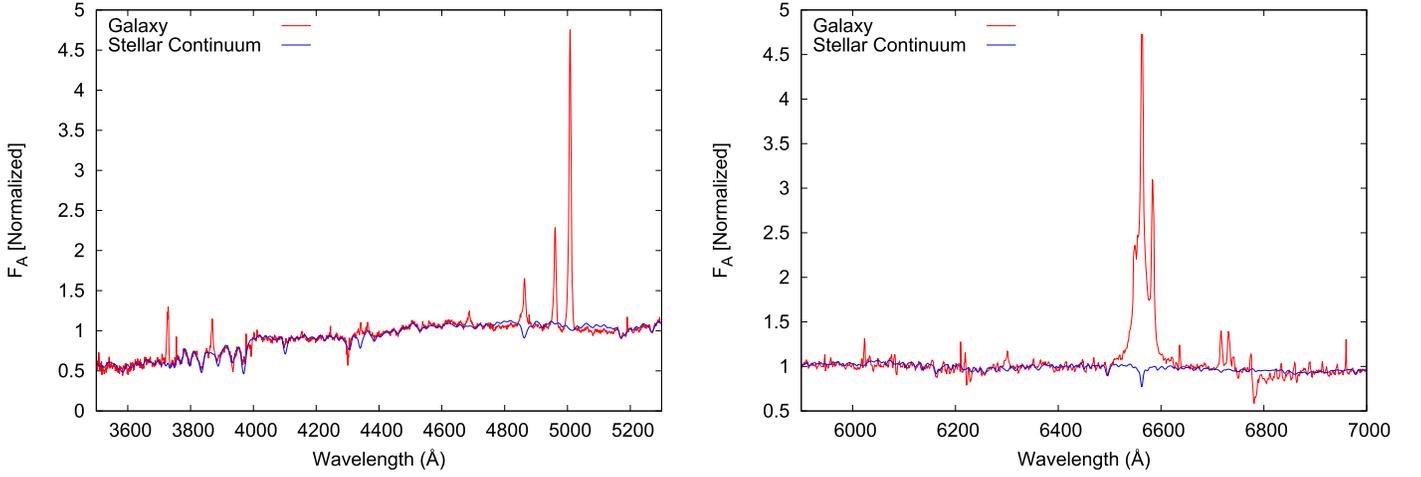

**Figure 2.** Stellar synthesis demonstration with Mrk 766. The galaxy spectra are plotted in red curves, while the synthesized stellar continuum in blue. On the *x*-axes are the wavelengths and on the *y*-axes are the fluxes normalized with respect to the continuum.

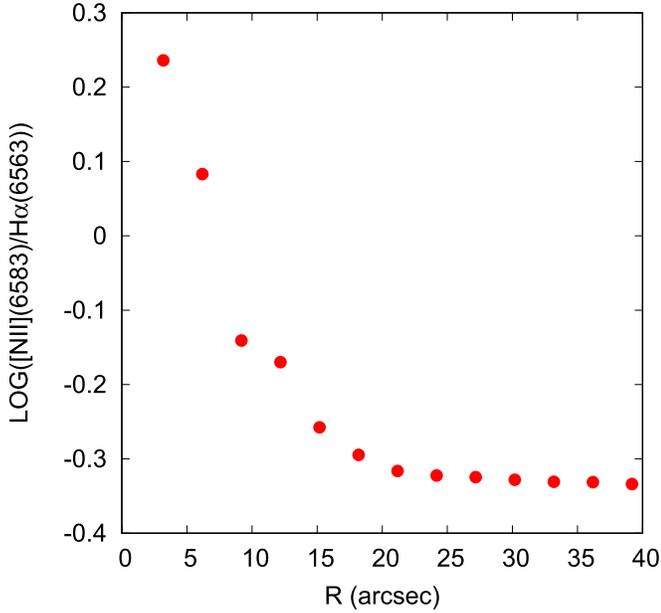

**Figure 3.** Decreasing trend of [N II]/H$\alpha$ with respect to radial distances from the center of NGC 1667.

we calculated the extinction factor by Equation (1):

$$f_{\rm int}(\lambda) = f_{\rm obs}(\lambda)\, 10^{0.4 A_\lambda}, \quad (1)$$

where $f_{\rm int}$ and $f_{\rm obs}$ are the intrinsic and observed fluxes respectively. The extinction $A_\lambda$ at wavelength $\lambda$ is related to the color excess $E(B-V)$ and the reddening curve $k(\lambda)$ derived in Calzetti et al. (2000) by

$$A_\lambda = k(\lambda) E(B-V). \quad (2)$$

The color excess $E(B-V)$ was calculated with H$\alpha$/H$\beta$ Balmer decrement as described by Momcheva et al. (2012):

$$E(B-V) = \frac{2.5}{k({\rm H}\beta) - k({\rm H}\alpha)} \log\left(\frac{({\rm H}\alpha/{\rm H}\beta)_{\rm obs}}{({\rm H}\alpha/{\rm H}\beta)_{\rm int}}\right) \quad (3)$$

in which we assumed $({\rm H}\alpha/{\rm H}\beta)_{\rm int}$ to be 2.86 in Case B recombination from Osterbrock (1989) where electron density is $10^{-4}\,{\rm cm}^2$ and temperature is 10,000 K.

We separated our 24 observed galaxies into two groups. In Figure 5, we present the emission line ratios measured from Seyfert 2s in circles and squares and LINERs in triangles. Meanwhile, in Figure 6, Seyfert 1s are plotted with circles and squares, star-forming-AGN composite objects with triangles, and star-forming spirals with diamonds. The averages of Seyfert spectra in nuclear and total apertures are presented, respectively, in Figures 5 and 6 with green pentagons. However, we were not able to measure the [S II] emission of Mrk 463. We have assumed that Mrk 463 does not change classification in the [S II]–[O III] diagnostic based on the consistent result in the other three BPT diagrams. We adopted the nuclear activity classification type of most of our sample galaxies from the NASA/IPAC Extragalactic Database that agreed with our measurements. However, we observed two galaxies with ambiguous spectroscopic classification—NGC 1144 and NGC 5953. The former's nucleus lies close to the AGN/normal galaxy boundaries in the [N II]–[O III], [S II]–[O III], and [O I]–[O III] planes but in the star-forming region of the [O I]–[O II] plane. The latter is located in the the Seyfert region in the [N II]–[O III] diagram but inside the star-forming boundary lines of all the others. Thus we decided the classification of these two galaxies by the majority of classification diagrams, i.e., NGC 1144 is classified as a Seyfert 2 but NGC 5953 as a star-forming/AGN composite.[4] The four BPT diagrams all show the effect of changing aperture size: the arrows point from the nuclei to the integrated host galaxies. To visualize the aperture effect, we kept the nuclear activity classification types for the integrated spectra so that it would be clear to see the population of AGNs that changed to different classification types in large apertures.

The nucleus-to-total shift for a given galaxy is not always in the same direction in all four BPT diagrams, particularly for those made with [O I]6300 and [O II]3727. Some galaxies shift upward in the BPT diagrams when the integrated spectra were measured. The same phenomenon was mentioned by Maragkoudakis et al. (2014), who also have occasionally observed enhancements of AGN features in larger apertures. One unlikely possibility is that metal-poor H II regions in the outer parts of the host galaxy produce the extra [O III] emission. However, this is inconsistent with three of our Seyfert 1s: FSC 04385-0828, NGC 3516, and Mrk 704. These three galaxies have an increased [N II]/H$\alpha$, but decreased [O III]/H$\beta$ in the largest apertures. Furthermore, unlike Seyfert 2s, Seyfert 1s in Figure 6 show a trend when the galaxies are observed in large (or total) apertures—the largest drop with increasing aperture is in the [O III]/H$\beta$ ratio (vertical axis). In contrast, the [N II]/H$\alpha$ and [S II]/H$\alpha$ ratios (horizontal axis) do not change much on average from the nucleus to the total aperture. These differences indicate that much of the extended line emission in the host galaxies of the Seyfert 1 nuclei is not from classic H II regions. Instead, there is a strong contribution from "diffuse ionized gas" (DIG). This gas can make up a substantial fraction of the total H$\alpha$ in a spiral galaxy (Zurita et al. 2000). It may be produced by photoionization with a diluted ionizing source, or by shocks (Hong et al. 2013). In either case, this DIG emission has relatively weak [O III], but strong [N II] and [S II] lines that are not very different from those emitted by the Seyfert NLR. This is why the measurement of the [O III]/H$\beta$ ratio is particularly important in separating out the emission contributed from the Seyfert nucleus.

---

[4] The NASA/IPAC Extragalactic Database (NED) is operated by the Jet Propulsion Laboratory, California Institute of Technology, under contract with the National Aeronautics and Space Administration.

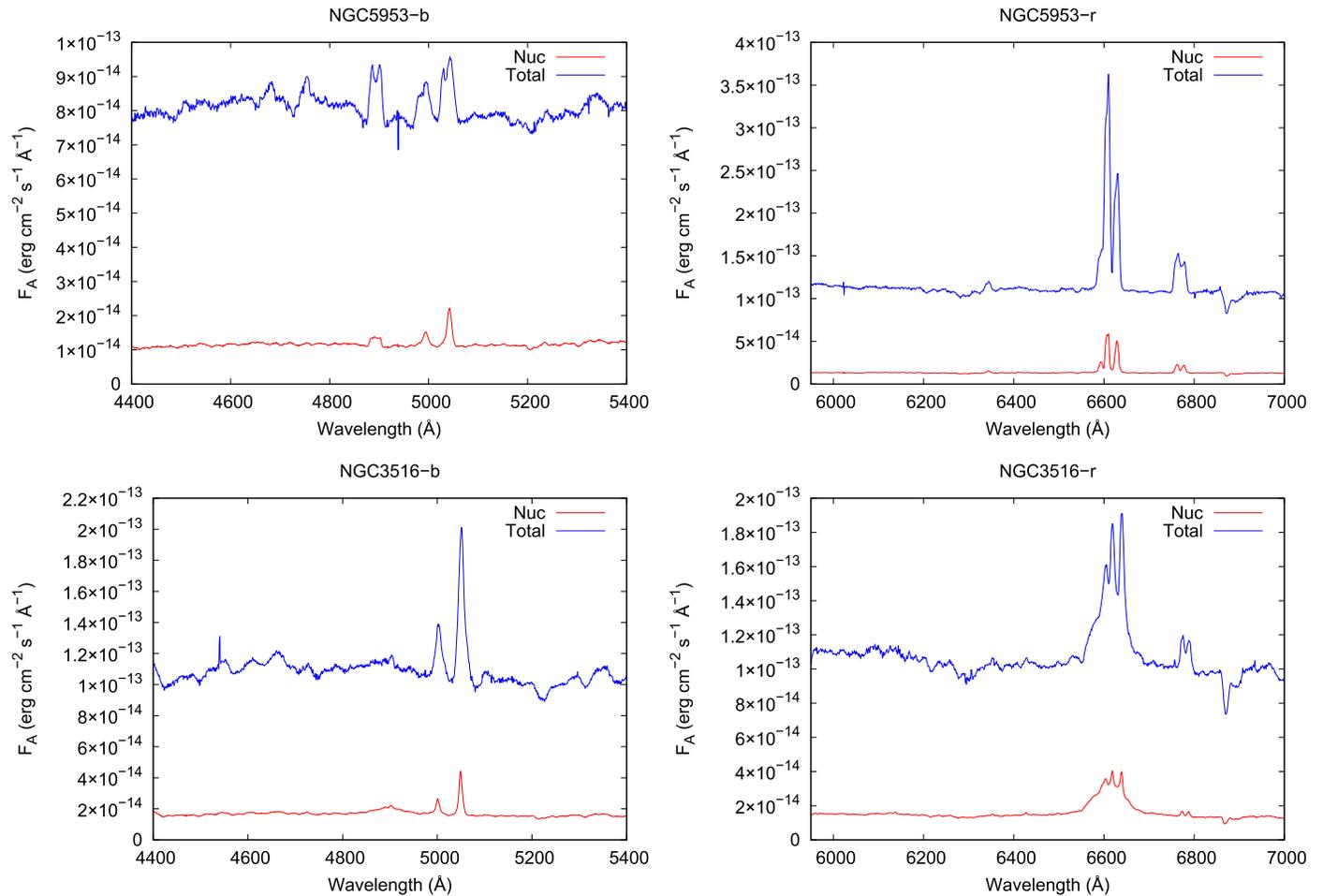

**Figure 4.** Spectra (without redshift correction) of NGC 5953 and NGC 3516. In each panel, the integrated spectra (top curves) and nuclear spectra (bottom curves) are plotted. In the left two panels are [O III] and H$\beta$ emission lines. In the right two panels are the [N II], [S II], and H$\alpha$ emission lines.

Our interpretation has assumed that the AGN-dominated line emission is entirely concentrated into our central "nuclear" pixel (typically 3 by 5 $\sim$ 12 arcsec). The NLR emission in some luminous AGNs and quasars has been found to be substantially extended—an "extended NLR" (ENLR) detected out to $\sim$10 kpc from the galactic center (Bennert et al. 2006a, 2006b; Hainline et al. 2013). Unambiguous evidence of an ENLR would, for example, be a consistently reversed gradient, i.e., in all BPT diagrams the total emission line ratio (blue box) could lie above and to the right of the nucleus emission (red circle). Although we cannot rule out the possibility, none of our observed sample of AGNs shows clear evidence of having a bright ENLR. This is not surprising, however, because most previously discovered ENLRs have only been found in AGNs that are >2 orders of magnitude more luminous than the AGNs in our study.

### 3.2. Spectroscopic Classification Statistics

Disagreements exist among the four BPT diagrams. On the other hand, the ambiguity of the boundary defined by Kewley et al. (2001) between AGNs and star-forming galaxies has already been pointed out by Stasinska et al. (2006), who estimated that this theoretical boundary may have overestimated the number of star-forming galaxies by 20%. But these ambiguities can hardly affect our average statistical results. Compared to the Seyfert 1s in Figure 6, Seyfert 2s show a larger probability of changing activity classification type when the aperture cover increases to the whole galaxy. In the first panel of Figure 5 where the diagnostic is done in the [N II]–[O III] plane, three Seyfert 2s have shifted to the composite AGN-star-forming region between Kewley et al. (2001) and Kauffmann et al. (2003). The other three BPT diagrams in Figure 5 show a consistent trend. In contrast, only one Seyfert 1—NGC 3982—has changed classification type in Figure 6. The broad wings of this Seyfert 1 are not quite visible in the large aperture, which is in agreement with the previous study that in rare cases some low-luminosity Seyfert 1s may lose their broad wing features due to the dilution of strong host galaxy continuum (Eun et al. 2017). Overall, NGC 1667 and NGC 3982 have consistently changed classification types throughout all four BPT diagrams when the aperture size grows to cover the whole host galaxy. NGC 1241 and IC 5298 have changed classification in three of the four BPT diagrams and both have migrated very close to the H II/Composite region in the remaining diagram respectively. NGC 1144 has changed its classification type in the BPT diagrams of [S II]–[O III] and [O I]–[O II], while Mrk6 only has changed classification in the [N II]–[O III] diagram.

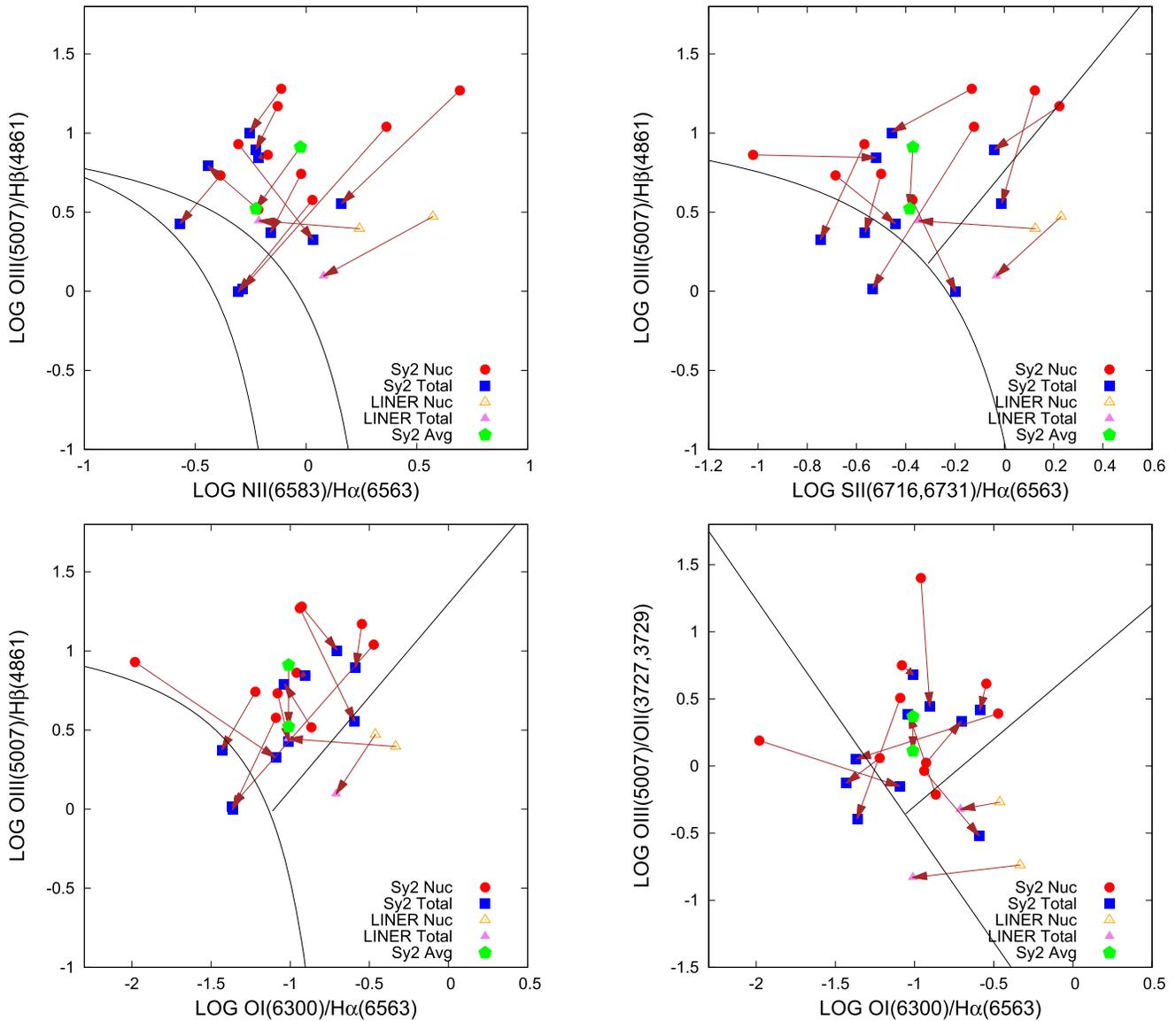

**Figure 5.** Emission line ratio diagrams of Seyfert 2s and LINERs. The black curves from Kewley et al. (2001, 2006) and Kauffmann et al. (2003) separate star-forming galaxies from AGNs. Different types of AGNs are represented with different symbols, explained in the legend.

In Figure 6, we have four galaxies—Mrk 1034, NGC 1056, NGC 5257, and NGC 5953—in the H II/Composite regions of the BPT diagrams. Mrk 1034 turns out to be a star-forming galaxy as it consistently appears below the demarcation curves defined by Kauffmann et al. (2003) and Kewley et al. (2001, 2006). NGC 1056, NGC 5257, and NGC 5953 are classified as composite AGN-star-forming galaxies based on all four BPT diagrams. In the [N II]–[O III] diagnostic, NGC 5953 has migrated from an AGN to H II region, and NGC 5257 has shown the migration caused by aperture effect in the [S II]–[O III] and [O I]–[O III] diagnostics.

It is also noticeable that some of our observed Seyfert galaxies migrate toward the LINER regions with varied apertures. The boundaries in the BPT diagrams defined in Kewley et al. (2001, 2006) were questioned by Stasinska et al. (2008) and Cid Fernandes et al. (2010). It has been shown that the division between Seyferts and LINERs may be blurred if the lines are very weak and not measured with high S/N (Cid Fernandes et al. 2010). This may further imply that contamination from the host galaxy could also shift a Seyfert to a LINER. However, in the rest of this paper, we will focus

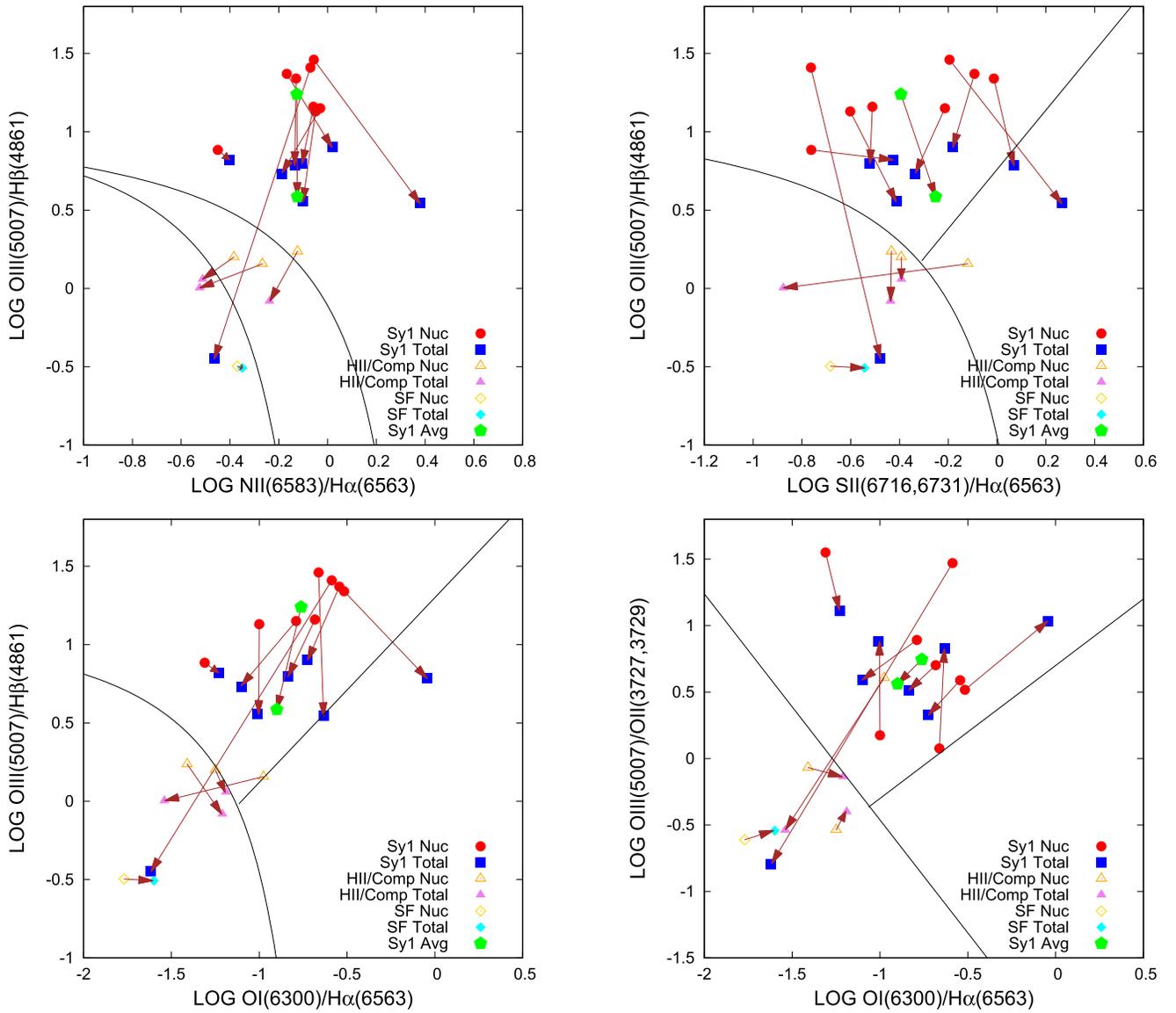

**Figure 6.** Same BPT emission line ratio classification diagrams as in Figure 4, except for Seyfert 1s and star-forming galaxies.

Table 2
Averages of Different Galaxy Types

| | | [N II]/Hα | [O III]/Hβ | [S II]/Hα | [O III]/[O II] | [O I]/Hα |
|---|---|---|---|---|---|---|
| Seyfert 1 | Nuc | −0.13 | 1.24 | −0.39 | 0.75 | −0.76 |
| | Total | −0.12 | 0.59 | −0.25 | 0.56 | −0.90 |
| Seyfert 2 | Nuc | −0.03 | 0.91 | −0.37 | 0.37 | −1.01 |
| | Total | −0.23 | 0.52 | −0.38 | 0.11 | −1.01 |

**Note.** The coordinates of the ensemble of galaxies in each BPT diagram as a function of nuclear spectrum type. The line ratios are given as logarithms.

on the first case because it represents the majority of potential misclassification.

By taking the average of the *x* and *y* coordinates of each Seyfert in BPT diagrams, we found a trend that the ensemble of the Seyfert 1s and 2s migrate toward the curves (Kewley et al. 2001, 2006) separating AGNs and star-forming galaxies. Out of the 18 Seyferts, those showing changed activity classification type are only a small portion but are consistent with all four diagnostics. We present the statistics in Tables 2 and 3. On average, four Seyfert galaxies from our sample would be

Table 3
Number of Changes in Larger Aperture

| | H II and Composite | Seyfert | LINER |
|---|---|---|---|
| Nuc | 4 | 18 | 2 |
| Total | 8 | 15 | 1 |

**Note.** Statistics of the aperture-changing activity classification type. The first row represents the number of galaxies classified according to the nuclear emission line ratio. The second row shows the classification according to emission lines from the whole galaxy.

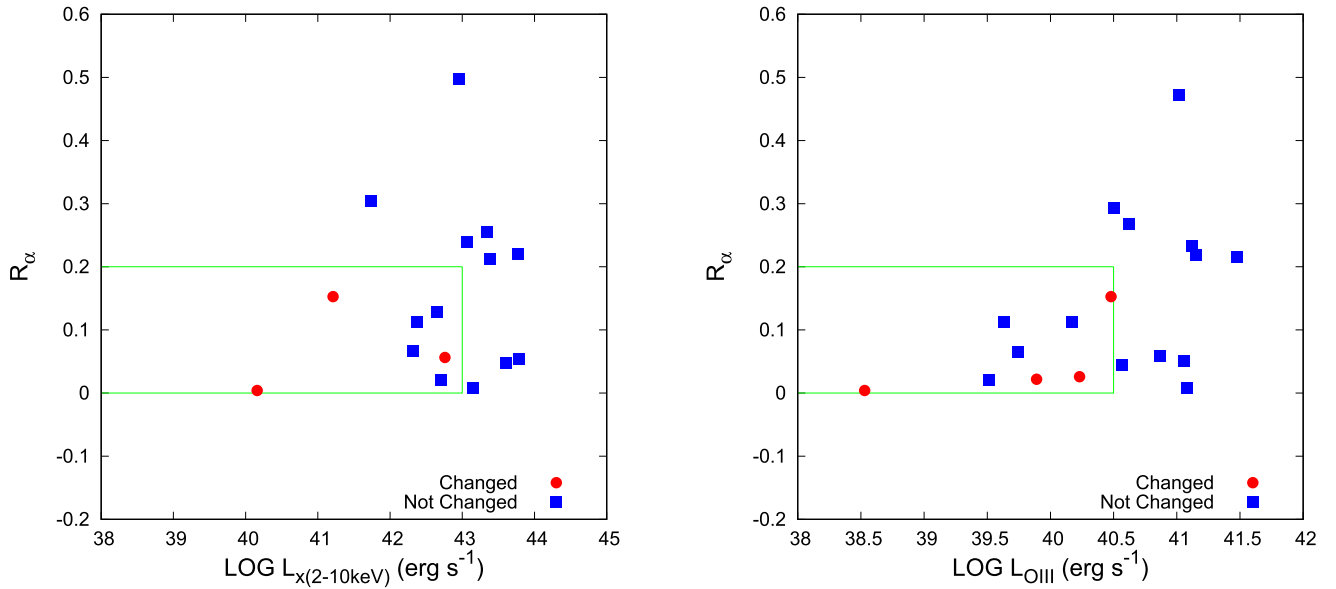

**Figure 7.** In the left panel on the *x*-axis is the nucleus hard X-ray (2–10 keV) luminosity of our Seyfert galaxies. In the right panel, the *x*-axis is the nuclear [O III] emission line luminosity. On *y*-axes is the nucleus to total Hα flux ratio. Those "missed" Seyferts that have changed classification type are plotted in red circles, while those have not are in blue squares.

misclassifed in at least three BPT diagrams with varied aperture sizes. This is only 22.2% of our Seyfert samples.

### 3.3. Galactic Luminosity and Its Correlation with Misclassification

Cowie et al. (2003) found that broad-line objects (Seyfert 1s) make up the majority of the luminous hard X-ray sources with $L_x > 10^{44}$ erg s$^{-1}$ at both $z = 2 \sim 4$ and $z = 0.1 \sim 1$. The intermediate hard X-ray sources with $L_x \sim 10^{42-43}$ erg s$^{-1}$ are powered by a mixture of obscured accretion onto super massive black holes and "normal" galaxies. Meanwhile, Barger et al. (2002) found that the 2–8 keV hard X-ray luminosity of star-forming galaxies never exceeds $10^{42}$ erg s$^{-1}$ based on the study of galaxies at $z = 0.84 \sim 1.02$. We acquired the nuclear 2–10 keV obscuration-corrected $L_x$ of 17 of our Seyfert 1s and 2s from Polletta et al. (1996), Brightman & Nandra (2011), and Nuria (2015). We also calculated the nuclear [O III] luminosity of 18 Seyferts in our sample based on our observed spectra considering the close relation between hard X-ray and [O III] luminosity, as shown by Malkan et al. (2017).

We calculated the nuclear fractions of our Hα emission lines according to

$$R_\alpha = \frac{F_{\text{nuc}-3}}{F_{\text{total}}}, \quad (4)$$

where $F_{\text{nuc}-3}$ stands for the fluxes from our nuclei (3″ subaperture) and $F_{\text{total}}$ stands for the integrated fluxes of the whole host galaxy. $R_\alpha$ represents the nuclear fraction of the $H_\alpha$ emission. In Figure 7, the results of Equation (4) are plotted against $L_x$ and $L_{\text{[O III]}}$. We also draw two boxes enclosing the Seyferts that have changed activity classification in our analysis. In both panels of Figure 7, we can see that about 43 ~ 50% Seyferts inside the green boxes could change their classification in the large aperture covering the whole host galaxy. We could not find the X-ray luminosity for NGC 1241 and Mrk 817. Since NGC 1241 has changed its classification due to aperture effect, the two panels in Figure 7 may look even more alike if we had those two objects in the hard X-ray plot. This result may help to answer the question raised by Barger et al. (2001a, 2001b) as mentioned in Section 1. The limits below which a large portion (up to 50%) of Seyfert galaxies may change its classification may therefore be

$$\log L_{x(2-10)} \text{ (erg s}^{-1}\text{)} < 43$$
$$\log L_{\text{[O III]}} \text{ (erg s}^{-1}\text{)} < 40.5$$
$$R_\alpha < 0.2.$$

We note that all of the galaxies observed in this study, as well as the spectral classification lines we have adopted in the BPT diagrams, are all at low redshifts. However, at redshifts $z \sim 2$, the boundary lines between Seyferts, star-forming, and composite galaxies appear to be shifted upward in Juneau et al. (2014). This may be caused by the lower metallicity of gas and O stars at $z \sim 2$ (Kewley et al. 2013a, 2013b). Meanwhile, Juneau et al. (2014) also noticed that the emission from H II regions photoionized by young stars shows higher [O III]/Hβ ratios than in the current ($z < 0.1$) epoch. Thus our results about Seyfert galaxy misclassification at low redshifts may not be strictly applicable to Seyfert galaxies observed at high redshifts. What we can say—qualitatively—is that the observational separation between Seyferts and non-Seyferts in the primary BPT diagram (with [O III]/Hβ on its vertical axis) becomes smaller at high redshift. In the presence of realistic uncertainties (because the spectra are sometimes noisy), a somewhat higher rate of spectral misclassification might be expected, regardless of what aperture size is used for the observations. These classification errors could go either way (star-forming misclassified as AGNs, or AGNs misclassified as star-forming). In addition to these random observational errors, the systematic observational errors that we have measured when large apertures are used, will also be present. We cannot say with confidence whether these systematic misclassifications would be worse than what we find at low redshifts, or not. Two countervailing trends will be at work. On the one hand, spectral classifications will tend to be more accurate because the nuclear luminosities of high-redshift AGNs tend to be higher. But, on the other hand, the host galaxies of high-redshift AGNs will also tend to be more luminous than in the current epoch. If the brighter starlight from the host galaxy becomes relatively more dominant, the aperture misclassification effect we find could actually become somewhat worse. At high redshifts ($z > 2$) the misclassification problem may be further addressed by including observations of diagnostic emission and absorption lines observable in the rest-frame ultraviolet (Hainline et al. 2011).

### 3.4. Comparison with Previous Results

Moran et al. (2002) obtained a similar result to ours. Among our 24 targets, NGC 262 (=Mrk 348), NGC 1667, and NGC 3982 overlap with the sample presented by them. We both found that NGC 262 does not change activity classification type, while the others do. They predicted that 60% of the Seyfert 2s would not be correctly classified based on the result that 11 galaxies out of their 18 targets have changed classification in the integrated spectra. Their samples consisted of 14 objects with $B$-band luminosity less than $10^{43}$ erg s$^{-1}$ and 4 with $L_B < 2 \times 10^{43}$ erg s$^{-1}$. In Barger et al. (2002), a linear correlation between the $L_B$ and 2–8 keV $L_x$ was presented, where approximately $1 < \log(L_B/L_{2-8}) < 10$. Therefore, the sample objects of Moran et al. (2002) have $L_x$ all less than $2 \times 10^{42}$ erg s$^{-1}$ and most less than $10^{42}$ erg s$^{-1}$. In fact, besides the three AGNs mentioned before, we have found the $L_{2-10}$ of six other AGNs from Comastri (2005), Della Ceca et al. (2008), and Maiolino et al. (1998). All of those AGNs have the 2–10 keV hard X-ray luminosity below our proposed classification threshold of $10^{43}$ erg s$^{-1}$. Thus their findings could be consistent with ours.

Recently, Thomas et al. (2017) provided their analysis on the Siding Spring Southern Seyfert Spectroscopic Survey (S7). They extracted a 4″ circular region from the nucleus of each galaxy. Due to the large range of redshifts, the same aperture size covers 0.1 ∼ 1.9 kpc. They have observed that a substantial fraction of their S7 samples (∼30%) in this aperture size lie below the Kewley et al. (2001) curves in BPT diagrams due to starlight contamination. This is close to what we have found from our targets. The S7 samples have nuclear [O III] luminosities between $10^{38} \sim 10^{42}$ erg s$^{-1}$. It may indicate that a slightly larger fraction of their samples are below our proposed classification threshold of $10^{40.5}$ erg s$^{-1}$, then their findings could be consistent with ours.

In addition, Theios et al. (2016) predicted that ∼67% of Seyferts would be misclassified at $z \sim 0.3$, where the typical aperture size becomes comparable to the galaxy's diameter. They instead used the equivalent width of H$\alpha$ emission line as their proxy criterion to predict the activity classification reliability. As more of the host galaxy at higher redshift is included within the measurement aperture, the ratio of H$\alpha$ emission line flux from the nucleus to the extended H II regions will decrease. They conservatively assumed that as the aperture size increases, as soon as the nuclear H$\alpha$ flux drops to one-third of the total H$\alpha$ flux from the host galaxy, a Seyfert galaxy will be misclassified as a star-forming galaxy. But as shown in Figure 7, in our analysis most of the Seyferts with nuclear H$\alpha$ fraction lower than one-third still remained as AGNs in the large aperture. This difference is not a result of which lines we measured. Figure 8 shows that the nuclear fraction of [N II], H$\alpha$, and [O III] we measured are correlated in our spectral data

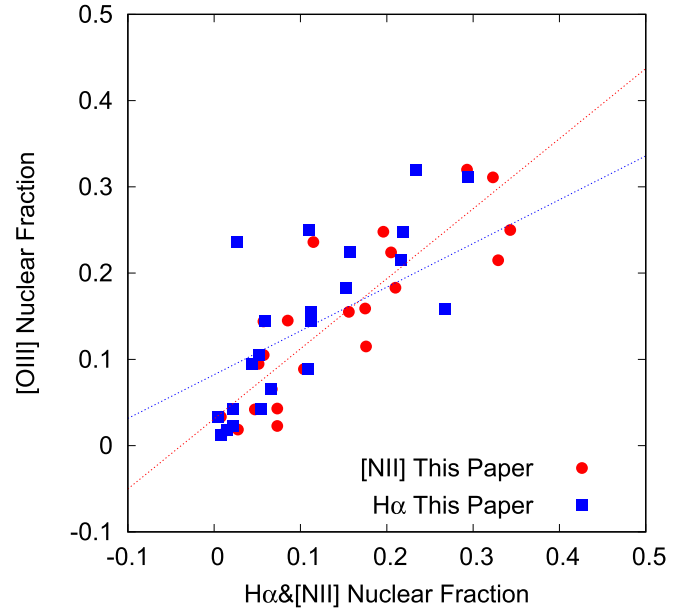

**Figure 8.** Vertical axis shows the nuclear fraction of total [O III] emission as measured from our spectral data cubes. The horizontal axis shows the nuclear fraction of H$\alpha$ and [N II]. The symbols are explained in the legend. The red and blue dotted lines are the least mean square fits of our calculated H$\alpha$ and [N II] nuclear fractions respectively.

cube—when one of these lines is strongly concentrated in the nucleus, all of them tend to be.

In the eight Seyferts common to our sample and theirs, only NGC 1241 and NGC 1667 have shifted classification in BPT diagrams to "composite/H II" region based on our spectroscopy. Less than 5% of H$\alpha$ emission of these three galaxies falls within our nuclear aperture. In contrast, NGC 1144, NGC 1194, NGC 1320, Mrk 79, NGC 262, and Arp 220, which have nuclear H$\alpha$ fractions smaller than one-third, did not change classification type in the BPT diagrams when their entire host galaxy light was included. Five out of those six galaxies (except NGC 1144) have either large nuclear H$\alpha$ fractions or luminosities beyond the range defined in Section 3.3. This indicates that only considering the fraction of nuclear H$\alpha$ emission (Theios et al. 2016) is not sufficient to determine whether a Seyfert will be misclassified.

To further investigate this difference, we were able to make a direct comparison of our spectroscopy versus their narrowband imaging. Our H$\alpha$+[N II] fluxes are larger than their data on average 0.2 ± 0.4 dex in nuclei and 0.4 ± 0.4 dex in integrated emission based on the 10 galaxies we both have observed. There is a substantial random scatter, of 0.4 dex, which could be attributable to the difficulty of using narrowband images to obtain accurate fluxes of low-surface brightness line emission. In the mean time, we have confirmed that our absolute flux scales are reliable, by comparison with four of our galaxies that also have SDSS spectroscopy. Our nuclear emission line fluxes agree with the SDSS line fluxes (tabulated in SDSS Data Release 7 (DR7), Abazajian et al. 2009) to within ±10% with no systematic shifts. Taking the detection efficiency and uncertainties into account, we find that the result of Theios et al. (2016) is not very different from ours. If observations were made with the same sensitivity, both studies indicate Seyfert misclassification when $R_\alpha$ drops below 0.2.

## 4. Conclusion

From a comparison of nuclear and integrated galaxy spectra, we have found that the location of an active galaxy in the BPT diagrams shifts due to the dilution host galaxy emission. This shift occurs among all three regions of the BPT diagrams—AGN, LINER, and star-forming galaxy regions, but is most pronounced at the boundary between AGN and star-forming galaxies. For our ensemble of Seyferts, the BPT line ratios will shift down toward the H II region, but the majority of activity classification type remains unchanged. The results throughout the BPT diagrams are consistent.

In our 18 Seyfert galaxies, only 22.2%—all with relatively less powerful active nucleus—would be misclassified at higher redshift. This portion would likely be as high as ~50% for low-luminosity objects, which are not well represented in our sample, but are more numerous in some other recent studies.

The case of Seyfert 1, NGC 3982, shows the possibility of being misclassified based on BPT diagrams and also because the broad wings of its H$\alpha$ line are lost due to starlight dilution. In contrast, the broad wing is substantially large in all other Seyfert 1s, which are correctly classified in large-aperture spectra. Therefore, the two different classification criteria—the existence of broad lines and the emission line ratios—may suffer the same result from the aperture effect.

At higher redshifts, the population of misclassified AGNs will depend on the population of low-luminosity objects. Cowie et al. (2003) found the number density of hard X-ray sources at $z = 0.1-1$ peaks at $L_x \sim 10^{41.7}$ erg s$^{-1}$. Thus it is still possible that a large number of Seyferts at $z > 0.3$ would be misclassified based on this population of low $L_x$ objects. The misclassification problem could be worse at higher redshift if the host galaxies of high-redshift AGNs also have substantially stronger star formation than the local ones we studied.

However, because the BPT diagrams may not be applicable to galaxies at high redshifts, more rest-frame UV features may be needed for reliable spectral classifications. Thus it is necessary to investigate methods other than the standard BPT diagrams.

This research has made use of the NASA/IPAC Extragalactic Database (NED), which is operated by the Jet Propulsion Laboratory, California Institute of Technology, under contract with the National Aeronautics and Space Administration.

### ORCID iDs

Junjie Xia  https://orcid.org/0000-0003-1412-092X
Matthew A. Malkan  https://orcid.org/0000-0001-6919-1237

## References

Abazajian, K. N., Adelman-McCarthy, J. K., Agüeros, M. A., et al. 2009, ApJS, 182, 543
Baldwin, J. A., Philips, M. M., & Terlevich, R. , 1981, PASP, 93, 5
Barger, A. J., Cowie, L. L., Brandt, W. N., et al. 2002, ApJ, 124, 1839
Barger, A. J., Cowie, L. L., Mushotzky, R. F., & Richards, E. A. 2001a, ApJ, 121, 662
Barger, A. J., Cowie, L. L., Bautz, M. W., et al. 2001b, ApJ, 122, 2177
Bennert, N., Treu, T., Auger, M. W., et al. 2015, ApJ, 809, 20
Bennert, V. N., Falcke, H., Schulz, H., Wilson, A. S., & Wills, B. J. 2002, ApJL, 574, L105
Bennert, V. N., Jungwiert, B., Komossa, S., Haas, M., & Chini, R. 2006a, A&A, 446, 919
Bennert, V. N., Jungwiert, B., Komossa, S., Haas, M., & Chini, R. 2006b, A&A, 456, 953
Blanton, M. R., Bershady, M. A., Abolfathi, B., et al. 2017, AJ, 154, 28
Brightman, M., & Nandra, K. 2011, MNRAS, 414, 3084
Bundy, K., Bershady, M. A., Law, D. R., et al. 2015, ApJ, 798, 7
Calzetti, D., Armus, L., Bohlin, R. C., et al. 2000, ApJ, 533, 682
Catalán-Torrecilla, C., Gil de Paz, A., Castillo-Morales, A., et al. 2015, A&A, 584, A87
Cid Fernandes, R., Mateus, A., Sodré, L., Stasińska, G., & Gomes, J. M. 2005, MNRAS, 358, 363
Cid Fernandes, R., Stasinska, G., Schlickmann, M. S., et al. 2010, MNRAS, 403, 1036
Comastri, A. 2005, in Supermassive Black Holes in the Distant Universe, ed. A. J. Barger (Heidelberg: Springer), 242
Cowie, L. L., Barger, A. J., Bautz, M. W., Brandt, W. N., & Garmire, G. P. 2003, ApJL, 584, L57
Croom, S. M., Lawrence, J. S., Bland-Hawthorn, J., et al. 2012, MNRAS, 421, 872
Della Ceca, R., Severgnini, P., Caccianiga, A., et al. 2008, MmSAI, 79, 65
Edelson, R., & Malkan, M. 2012, ApJ, 751, 52
Eisenstein, D. J., Weinberg, D. H., Agol, E., et al. 2011, AJ, 142, 72
Eun, D., Woo, J., & Bae, H. 2017, ApJ, 842, 5
Green, A. W., Glazebrook, K., Gilbank, D. G., et al. 2017, MNRAS, 470, 639
Greene, J. E., Zakamska, N. L., Ho, L. C., & Barth, A. J. 2011, ApJ, 732, 9
Hainline, K. N., Hickox, R., Greene, J. E., Myers, A. D., & Zakamska, N. L. 2013, ApJ, 774, 142
Hainline, K. N., Shapley, A. E., Greene, J. E., & Steidel, C. C. 2011, ApJ, 733, 12
Heckman, T. M. 1980, A&A, 87, 152
Ho, L. C., Filippenko, A. V., Sargent, W. L. W., & Peng, C. Y. 1997, ApJS, 112, 391
Hong, S., Calzetti, D., Gallagher, J. S., III, et al. 2013, ApJ, 777, 63
Husemann, B., Jahnke, K., Sanchez, S. F., et al. 2013b, A&A, 549, A87
Husemann, B., Wisotzki, L., Sánchez, S. F., & Jahnke, K. 2013a, A&A, 549, A43
Iglesias-Páramo, J., Vilchez, J. M., Galbany, L., et al. 2013, A&A, 553, L7
Iglesias-Páramo, J., Vilchez, J. M., Rosales-Ortega, F. F., et al. 2016, ApJ, 826, 71
Juneau, S., Bournaud, F., Charlot, S., et al. 2014, ApJ, 788, 88
Kauffmann, G., Heckman, T. M., Tremonti, C., et al. 2003, MNRAS, 346, 1055
Kewley, L. J., Dopita, M. A., Leitherer, C., et al. 2013a, ApJ, 774, 100
Kewley, L. J., Dopita, M. A., Sutherland, R. S., Heisler, C. A., & Trevena, J. 2001, ApJ, 556, 121
Kewley, L. J., Groves, B., Kauffmann, G., & Heckman, T. 2006, MNRAS, 372, 961
Kewley, L. J., Maier, C., Yabe, K., et al. 2013b, ApJL, 774, L10
Liu, G., Zakamska, N. L., Greene, J. E., Nesvadba, N. P. H., & Liu, X. 2013a, MNRAS, 430, 2327
Liu, G., Zakamska, N. L., Greene, J. E., Nesvadba, N. P. H., & Liu, X. 2013b, MNRAS, 436, 2576
Maiolino, R., Salvati, M., Bassani, L., et al. 1998, A&A, 338, 781
Malkan, M. A. 1988, AdSpR, 8, 49
Malkan, M. A., & Filippenko, A. V. 1983, ApJ, 275, 477
Malkan, M. A., Jensen, L. D., Rodriguez, D. R., Spinoglio, L., & Rush, B. 2017, ApJ, 846, 102
Malkan, M. A., & Oke, J. B. 1983, ApJ, 265, 92
Maragkoudakis, A., Zezas, A., Ashby, M. L. N., & Willner, S. P. 2014, MNRAS, 441, 2296
Mast, D., Rosales-Ortega, F. F., Sanchez, S. F., et al. 2014, A&A, 561, A129
Momcheva, I., Lee, J. C., Ly, C., et al. 2012, arXiv:1207.5479
Moran, E. C., Filippenko, A. V., & Chornock, R. 2002, ApJL, 579, L71
Nuria, T. A. 2015, Master thesis, Universitat de Barcelona
Osterbrock, D. 1989, Astrophysics of Gaseous Nebulae and Active Galactic Nuclei (Mill Valley, CA: Univ. Sci. Books)
Polletta, M., Bassani, L., Malaguti, G., Palumbo, G. G. C., & Caroli, E. 1996, ApJS, 106, 399
Richards, E. E., Van Zee, L., Barnes, K. L., et al. 2016a, MNRAS, 460, 689
Richards, S. N., Bryant, J. J., Croom, S. M., et al. 2016b, MNRAS, 455, 2826
Rush, B., Malkan, M. A., & Spinoglio, L. 1993, ApJS, 89, 1
Sanchez, S. F., Kennicutt, R. C., Gil de Paz, A., et al. 2012, A&A, 538, A8
Spinoglio, L., & Malkan, M. A. 1992, ApJ, 399, 504
Stasinska, G., Cid Fernandes, R., Mateus, A., Sodré, L., Jr, & Asari, N. V. 2006, MNRAS, 371, 972

Stasinska, G., Vale Asari, N., Cid Fernandes, R., et al. 2008, MNRAS, 391, L29
Sun, A., Greene, J. E., Zakamska, N. L., et al. 2018, MNRAS, 480, 2302
Theios, R., Malkan, M. A., & Ross, N. R. 2016, ApJ, 822, 45
Thomas, A. D., Dopita, M. A., Shastri, P., et al. 2017, ApJS, 232, 11
Villar-Martin, M., Arribas, S., Emonts, B., et al. 2010, MNRAS, 460, 130
Wandel, A., Peterson, B. M., & Malkan, M. A. 1999, ApJ, 526, 579
Woo, J., Treu, T., Malkan, M. A., & Blandford, R. D. 2008, in Proc. IAU Symp. 245, Co-evolution of Bulges and Black Holes, ed. M. Bureau, E. Athanssoula, & B. Barbuy (Cambridge: Cambridge Univ. Press), 223
York, D. G., Adelman, J., Anderson, J. E., et al. 2000, AJ, 120, 1579
Zurita, A., Rozas, M., & Beckman, J. E. 2000, A&A, 363, 9